\documentclass[openacc]{rsproca_new}
\newcommand{\SSC}{S/S_{c}}
\newcommand{\Prm}{\mbox{\textit{Pm}}}

\newcommand\Beq{\begin{eqnarray}} 
\newcommand\Eeq{\end{eqnarray}}

\newcommand{\eq}[1]{eq.~(\ref{#1})}

\newcommand{\pd}[1]{\partial_{#1}}

\newcommand{\Reyn}{\mathrm{Re}}
\newcommand{\Reym}{\mathrm{Rm}}

\newcommand{\Ro}{\mathrm{Ro}}

\newcommand{\eps}{\varepsilon}

\newcommand{\s}{\text{s}}
\newcommand{\x}{\text{x}}
\newcommand{\y}{\text{y}}

\newcommand{\dd}[1]{\,\mathrm{d}{#1}}

\usepackage{graphicx}
\usepackage{amsmath}

\begin{document}

\title{The Magnetorotational Instability Prefers Three Dimensions}

\author{
Jeffrey~S.~Oishi$^{1}$, Geoffrey~M.~Vasil$^{2}$, Morgan Baxter$^{1}$,
Andrew Swan$^{3}$, Keaton~J.~Burns$^{4,5}$, Daniel~Lecoanet$^6$, Benjamin~P.~Brown$^7$}

\address{$^{1}$Bates College, Lewiston, ME 04240, USA\\
$^{2}$University of Sydney, Sydney, NSW, Australia\\
$^{3}$University of Cambridge, Cambridge,
UK\\
$^{4}$Flatiron Institute, New York, NY 10010, USA\\
$^{5}$MIT, Cambridge, MA 02139, USA\\
$^{6}$Princeton University, Princeton, NJ 08544, USA\\
$^{7}$University of Colorado, CO 80309, USA}
\corres{Jeffrey~S.~Oishi\\
\email{joishi@bates.edu}}

\begin{abstract}
The magnetorotational instability (MRI) occurs when a weak magnetic field destabilises a rotating, electrically conducting fluid with inwardly increasing angular velocity.
The MRI is essential to astrophysical disk theory where the shear is typically Keplerian.
Internal shear layers in stars may also be MRI unstable, and they take a wide range of profiles, including near-critical.
We show that the fastest growing modes of an ideal magnetofluid are three-dimensional provided the shear rate, $S$, is near the two-dimensional onset value, $S_c$.
For a Keplerian shear, three-dimensional modes are unstable above $S\approx0.10S_c$, and dominate the two-dimensional modes until $S\approx2.05S_{c}$.
These three-dimensional modes dominate for shear profiles relevant to stars and at magnetic Prandtl numbers relevant to liquid-metal laboratory experiments.
Significant numbers of rapidly growing three-dimensional modes remain well past $2.05S_{c}$. 
These finding are significant in three ways. 
First, weakly nonlinear theory suggests that the MRI saturates by pushing the shear rate to its critical value. 
This can happen for systems, like stars and laboratory experiments, that can rearrange their angular velocity profiles.
Second, the non-normal character and large transient growth of MRI modes should be important whenever three-dimensionality exists.
Finally, three-dimensional growth suggests direct dynamo action driven from the linear instability.
\end{abstract}

\maketitle

\section{Introduction}
The magnetorotational instability (MRI) is extremely important in astrophysical fluid dynamics.
A weak magnetic field catalyses turbulence in a Keplerian shear by changing the stability criterion for differentially rotating flows from a negative angular \emph{momentum} gradient to a negative angular \emph{velocity} gradient \cite{1998RvMP...70....1B,2010RSPTA.368.1607J}.
This discovery explained the ubiquitous accretion onto compact objects at rates compatible with observations, and may also influence the formation of planets \cite{2007Natur.448.1022J}.
In disks, the gravitational field dominates the local plasma dynamics, and thus the MRI cannot significantly affect the background shear; it must saturate by other means \cite{2018MNRAS.474.3451X}.
However, stars and liquid metal Taylor-Couette experiments have differential rotation profiles driven by much weaker stresses.
Where the MRI is active in these flows, it saturates by pushing the background shear close to critical \cite{2015RSPSA.47140699V,2017ApJ...841....1C,2017ApJ...841....2C}, analogous to convection mixing entropy.
Stellar interiors have extremely high fluid and magnetic Reynolds numbers, but can operate at or near the critical shear rate for the MRI.
This finite critical shear results from a finite-channel cutoff in the radial direction.
Despite the extensive literature on accretion disks (strong shear, low dissipation) and liquid-metal experiments (weak shear, large dissipation), even the linear MRI is not well understood in the weak-shear (e.g.\ $S \gtrsim S_c$), low-dissipation regime. 

Here, we investigate the stability of three-dimensional perturbations near the two-dimensional critical shear rate $S_{c}$ for a nearly inviscid, ideal MHD flow.
Throughout this paper, we use ``three-dimensional'' to refer to non-axisymmetric perturbations (and their local Cartesian equivalents), while ``two-dimensional'' refers to axisymmetric perturbations.
In both cases, we retain all three components of velocity and magnetic fields.
We find that the first destabilised modes are three-dimensional, and thus could act as a dynamo even in the absence of secondary instability.
These results also suggest that the non-normal growth of the MRI is always important, even when axisymmetric modes dominate.

\section{Methods}
\label{sec:methods}

We numerically solve the linearized magnetohydrodynamic equations in rotating plane Couette geometry.
This corresponds to a Cartesian frame rotating with angular frequency $\Omega$ and a linear background shear, $V_{y}=Sx$ \cite{2015RSPSA.47140699V}; it is also the narrow-gap limit of the Taylor-Couette geometry.
We cast the Navier-Stokes equation in the form,
\begin{equation}\label{eq:mhd}
\frac{D \boldsymbol{v}}{Dt}+f \boldsymbol{\hat{z}}\times\boldsymbol{v}+{S}v_{x}\,\boldsymbol{\hat{y}}+\boldsymbol{\nabla}{p}+\nu\,\boldsymbol{\nabla}\times\boldsymbol{\omega}=B_{0}\partial_{z}\boldsymbol{b},
\end{equation}
where
\begin{equation}
\boldsymbol{\omega}=\boldsymbol{\nabla}\times\boldsymbol{v},\quad\text{and}\quad\frac{D}{Dt}=\partial_{t}+{S}x\,\partial_{y}\end{equation}

We write the induction equation in terms of the $x$-component of the magnetic field,
\begin{equation}\label{eq:Bx}
\frac{Db_{x}}{Dt}+\eta(\partial_{y}j_{z}-\partial_{z}j_{y})=B_{0}\partial_{z}v_{x},
\end{equation}
and the $x$-component of the current density ($j_{x}=\partial_{y}b_{z}-\partial_{z}b_{y}$),
\begin{equation}\label{eq:Jx}
\frac{Dj_{x}}{Dt}-\eta\nabla^{2}j_{x}=B_{0}\partial_{z}\omega_{x}-S\partial_{z}b_{x}.
\end{equation}
We explicitly enforce divergence-free velocity and magnetic field,
\begin{equation}\label{eq:divu}
 \boldsymbol{\nabla}\cdot\boldsymbol{v}=\boldsymbol{\nabla}\cdot\boldsymbol{b}=0.
\end{equation}
The spatial domain is a doubly periodic channel in $y,\,z$ with width $-d/2\le{x}\le d/2$.
The boundary conditions are impenetrable stress-free and perfectly conducting; $v_{x}=\omega_{y}=\omega_{z}=b_{x}=\partial_{x}j_{x}=0$ at $x=\pm{d/2}$. 

The main input parameters are the Coriolis parameter, $f=2 \Omega$; the background shear rate, $S=dV_{y}/dx<0$; and the vertical magnetic field $B_{0}$ (in Alfv\'{e}n units $\mu_{0}\rho_{0}=1$).
Accretion-disk modelling usually considers the Rossby number $\Ro=-S/f<1$. 
The regime $\Ro\ge1$ corresponds to purely hydrodynamical Rayleigh unstable shear.
Unless otherwise stated $\Ro=3/4$ (Keplerian).
The solution also depends on the viscosity $\nu$ and resistivity $\eta$; in our non-dimensionalization, these are equivalent to the inverse Reynolds and Magnetic Reynolds numbers, respectively.
That is, $\Reyn = 1/\nu$ and $\Reym = 1/\eta$.

The MRI is a weak-field instability; in the inviscid, ideal case the critical shear rate for axisymmetric instability (i.e.\ in \emph{two dimensions}) is
\begin{equation}\label{eq:Sc}
  S_{c}=-\frac{\pi^{2}B_{0}^2}{fd^2}
\end{equation}
\cite{2015RSPSA.47140699V}.

We use $\SSC$ as our instability control parameter.
Because $\SSC$ serves as a ratio of the dimensionless magnetic field strength $B_0^2/f d$ to the length scale $d$, this serves to specify the background field strength.

We assume harmonic perturbations in $y$ and $z$, (e.g. pressure) $p=\hat{p}(x)e^{i(k_{y}y+k_{z}z)+\sigma{t}}$. 
We use a complex-valued growth rate $\sigma=\gamma+i\omega$ with $\gamma,\,\omega$ both real. 
The system reduces to a $\sigma$-eigenvalue problem of 10 first-order ODEs in $x$ with Dirichlet boundary conditions.
We pose and solve equations~(\ref{eq:mhd}-\ref{eq:divu}) using the \emph{Dedalus} framework \cite{2019arXiv190510388B}.

Our main interest is in ideal ($\eta=0$), inviscid ($\nu=0$) conditions.
However, we set $\eta=\nu=10^{-5}$ to avoid critical layers in the \textit{stable} solution branch.
We confirmed our results for unstable solutions are insensitive to small diffusion. 
For each $(k_{y},k_{z})$ pair, we solve the eigenvalue problem using $n_{x}=128$ modes; all our results are identical at double the resolution \footnote{See \protect\url{github.com/jsoishi/mri_prefers_3d} for all code used in this paper.}

For both the ideal and non-ideal MHD equations, we solve the eigenvalue problem in the $x$ direction using the \texttt{EigenValueProblem} solver in \emph{Dedalus} for a grid of $N_y \times N_z$ modes in the $y$ and $z$ directions.
For most of our runs, we use a targeted, sparse eigenvalue solver to find the 15 modes closest to a guess for the maximum growth rate.
In run 3, we have confirmed that dense solvers retrieve identical results.
For nearly ideal runs, we use the ideal 2D growth rate as input for the smallest $k_{y} > 0$ mode at each $k_{z}$ and then use the output from each previous $k_{y}$ as an input guess for the next mode.
For those runs that have significant $\eta$, we instead use a dense solve for the $k_{y} = 0$ modes and then step forward in $k_{y}$ at each $k_{z}$ as before.
Our solver is embarrassingly parallelised over the $k_{z}$ modes.
For the spectrum in figure 1 of the main text, we used a dense eigenvalue solver at $(k_{y}^{max},k_{z}^{max})$ for $\SSC= 1.002$.

In order to ensure our results are converged, we have repeated runs at $n_x=256$ Chebyshev modes as well as doubling the number of modes in $y$ and $z$.
\begin{table}
\begin{tabular}{cccccccccccccc}
\textbf{Run} & \textbf{$\SSC$} & \textbf{$\nu$} & \textbf{$\eta$} & \textbf{$\Ro$} & \textbf{$N_x$} & \textbf{$N_{k_{y}}$} & \textbf{$N_{k_{z}}$} & \textbf{Sparse/Dense}& \\
1  &   1.02 & $10^{-5}$ & $10^{-5}$ & 0.75 & 128 & 200 & 200 & sparse & resolution study\\
2  &   1.02 & $10^{-5}$ & $10^{-5}$ & 0.75 & 256 & 200 & 200 & sparse & \\
3  &   1.02 & $10^{-5}$ & $10^{-5}$ & 0.75 & 128 & 200 & 200 & dense  & \\
4  &   1.02 & $10^{-5}$ & $10^{-5}$ & 0.75 & 128 & 512 & 512 & sparse & \\
5  &   1.02 & $10^{-5}$ & $10^{-5}$ & 0.75 & 128 & 100 & 100 & sparse & \\
6  &   0.2  & $10^{-5}$ & $10^{-5}$ & 0.75 & 128 & 200 & 200 & sparse & $\SSC$ variation\\
7  &   0.3  & $10^{-5}$ & $10^{-5}$ & 0.75 & 128 & 200 & 200 & sparse & \\
8  &   0.4  & $10^{-5}$ & $10^{-5}$ & 0.75 & 128 & 200 & 200 & sparse & \\
9  &   0.5  & $10^{-5}$ & $10^{-5}$ & 0.75 & 128 & 200 & 200 & sparse & \\
10 &   0.64 & $10^{-5}$ & $10^{-5}$ & 0.75 & 128 & 200 & 200 & sparse & \\
11 & 1.002  & $10^{-5}$ & $10^{-5}$ & 0.75 & 128 & 200 & 200 & sparse & \\
12 & 1.002  & $10^{-5}$ & $10^{-5}$ & 0.75 & 128 & 200 & 200 & dense & \\
13 & 1.44   & $10^{-5}$ & $10^{-5}$ & 0.75 & 128 & 200 & 200 & sparse & \\
14 & 1.75   & $10^{-5}$ & $10^{-5}$ & 0.75 & 128 & 200 & 200 & sparse & \\
15 & 1.891  & $10^{-5}$ & $10^{-5}$ & 0.75 & 128 & 200 & 200 & sparse & \\
16 & 2.     & $10^{-5}$ & $10^{-5}$ & 0.75 & 128 & 200 & 200 & sparse & \\
17 & 2.01   & $10^{-5}$ & $10^{-5}$ & 0.75 & 128 & 200 & 200 & sparse & \\
18 & 2.015  & $10^{-5}$ & $10^{-5}$ & 0.75 & 128 & 200 & 200 & sparse & \\
19 & 2.031  & $10^{-5}$ & $10^{-5}$ & 0.75 & 128 & 200 & 200 & sparse & \\
20 & 2.05   & $10^{-5}$ & $10^{-5}$ & 0.75 & 128 & 200 & 200 & sparse & \\
21 & 2.1    & $10^{-5}$ & $10^{-5}$ & 0.75 & 128 & 200 & 200 & sparse & \\
22 & 2.25   & $10^{-5}$ & $10^{-5}$ & 0.75 & 128 & 200 & 200 & sparse & \\
23 & 2.5    & $10^{-5}$ & $10^{-5}$ & 0.75 & 128 & 200 & 200 & sparse & \\
24 & 4      & $10^{-5}$ & $10^{-5}$ & 0.75 & 128 & 200 & 200 & sparse & \\
25 & 1.02   & $10^{-6}$ & $10^{-6}$ & 0.75 & 128 & 200 & 200 & sparse & Reynolds number study\\
26 & 1.02   & $10^{-4}$ & $10^{-4}$ & 0.75 & 128 & 200 & 200 & sparse & \\
27 & 1.02   & $10^{-3}$ & $10^{-3}$ & 0.75 & 128 & 200 & 200 & sparse & \\
28 & 1.02   & $10^{-2}$ & $10^{-2}$ & 0.75 & 128 & 200 & 200 & sparse & \\
29 & 1.02   & $10^{-5}$ & $10^{-5}$ & 0.1  & 128 & 200 & 200 & sparse & Low Rossby\\ 
30 & 1.02   & $10^{-6}$ & $10^{-2}$ & 0.1  & 128 & 200 & 200 & sparse & Liquid metal case\\ 
\end{tabular}
\caption{Eigenvalue calculations performed.}
\label{tab:runs}
\end{table}

\begin{figure*}[ht]
  \includegraphics[width=\textwidth]{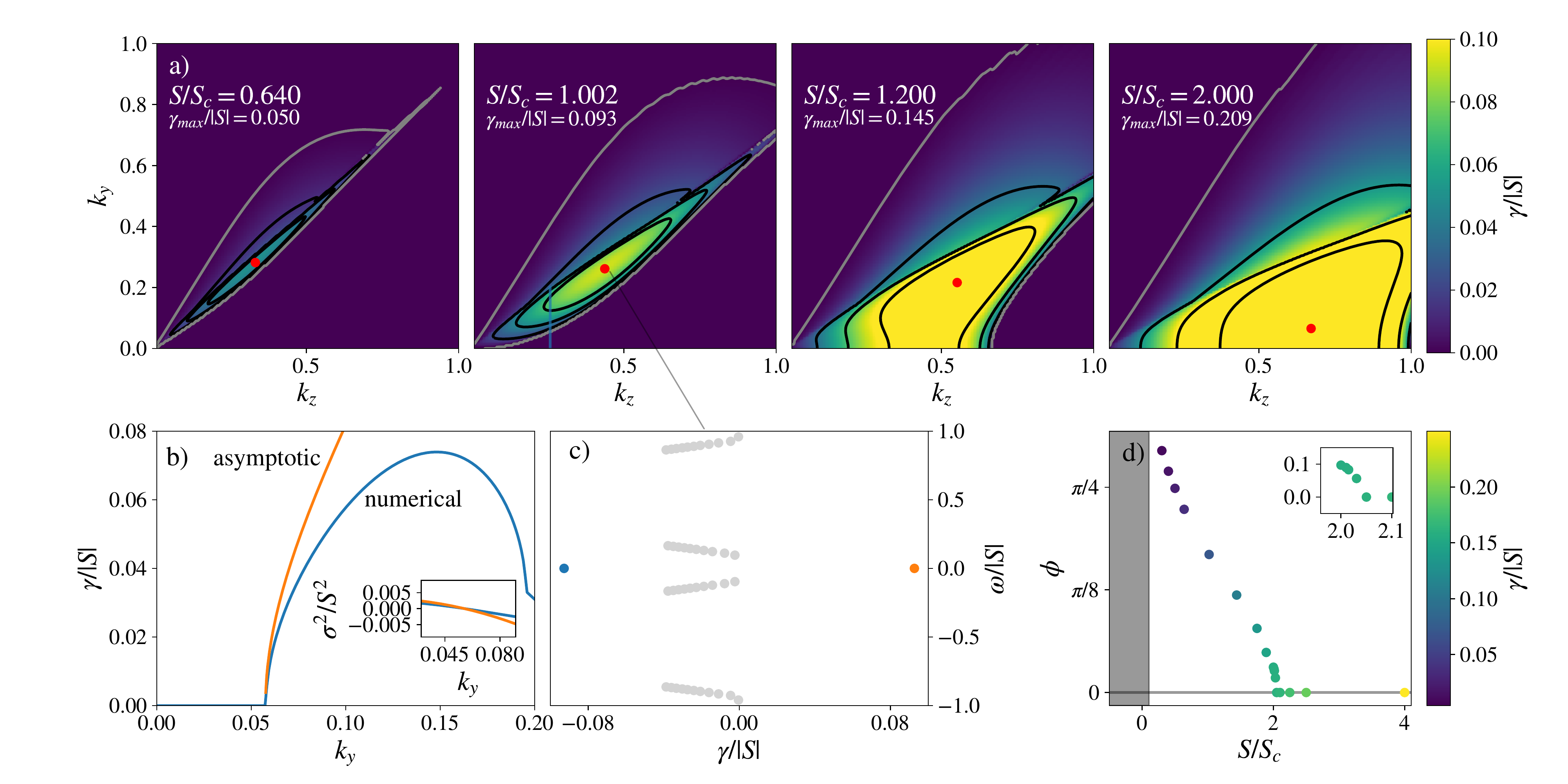}
  \caption{Growth rates for three-dimensional MRI modes. 
  a) growth rates $\gamma$ over a grid of $k_{y}$ and $k_{z}$ for four values of supercriticality $\SSC$. 
  The black contours are equally spaced between zero and the maximum growth rate.
  The grey contour highlights $\gamma=0$; the red dots indicate the fastest growing mode. 
Two dimensional modes occur at $k_{y} = 0$ on the bottom of each figure.
  At $\SSC=0.640$, there are no two dimensional modes. 
  b) Growth rate vs $k_{y}$ for $k_{z}=0.259$ at $\SSC=1.002$ (highlighted by the blue line in a). 
  The orange line gives the asymptotic result from equation~(\ref{eq:asymp}) and the blue line are the numerical results. 
  The asymptotic form is valid as $\SSC\to1$ and $k_{y}\ll1$.) c) The full discrete spectrum of the MRI for $(k_{y},k_{z})\simeq(0.263,0.447)$. 
  The unstable mode is plotted in orange, its stable complex conjugate is blue, and all other modes are grey. 
  d) The phase angle, $\arctan(k_{y}/k_{z})$, as a function of $\SSC$ showing the exact transition to purely two-dimensional fastest modes ($\phi=0$) at $\SSC\gtrsim2.05$. }
  \label{fig:growth_rate}
\end{figure*}

\section{Results}
\label{sec:results}

\subsection{Growth Rates and Three-dimensionality}
\label{sec:growth}

Our two main results are: 
(I) When the first two-dimensional mode becomes unstable, there already exist three-dimensional modes with positive growth rate. 
(II) At sufficiently large criticality, the fastest-growing mode becomes purely two-dimensional. 
These results are pertinent in two ways. 
First, the MRI contains a substantial ``Goldilocks regime'' with possible direct dynamo action, and this regime likely applies to stellar interiors and laboratory experiments. 
Second, our results accord with well-established results for accretion disks that expect two-dimensional primary modes.
Figure~\ref{fig:growth_rate}a shows the growth rates for four values of $\SSC$. 
At $\SSC=1.002$, the maximum growth rate ($\gamma_{\max}/|S|=0.093$) happens at $(k_{y},k_{z})\simeq(0.263,0.447)$.
Two-dimensional instability occurs for modes with finite growth rates at $k_{y} = 0$ along the bottom of the figure.
In all cases in figure~\ref{fig:growth_rate}a, the maximum growth rate does not occur along this line, indicating the dominance of three-dimensional modes.
The first panel of figure~\ref{fig:growth_rate}a shows $\SSC=0.640$, which contains only three-dimensional instability: the zero growth rate contour (highlighted in grey) does not intercept the $k_{z}$-axis.

\subsection{Asymptotic Calculation}
\label{sec:asymp}
The MRI's preference for three-dimensional modes can be predicted analytically.
We analyse equations~(\ref{eq:mhd}-\ref{eq:divu}) asymptotically close to $S_{c}$, and compute the leading-order correction to the growth rate from three-dimensional effects assuming $\SSC=1+\epsilon^{2}R$, $k_{z}\sim\mathcal{O}(\epsilon)$, $\sigma\sim{k_{y}}\sim\mathcal{O}(\epsilon^{2})$. 
Here, we outline the asymptotic calculation for the leading order correction to the 2D growth rates when accounting for three dimensional effects via finite $k_{y}$.
Being in the rotating frame means the shear has no net
\Beq
\int_{-d/2}^{d/2} V_{0,y}(x) \text{d} x  \ = \ 0 \quad \implies  \quad V_{0,y}(x) \ = \ S\, x
\Eeq
The full linear ideal equations are:
\Beq\label{u-eq}
\pd{t} v_{x} + S x \pd{y} v_{x} - f v_{y} + \pd{x} p - B_{0} \pd{z} b_{x} &=& 0\\
\pd{t} v_{y} + S x \pd{y} v_{y} + (f+S) v_{x} + \pd{y} p - B_{0} \pd{z} b_{y} &=& 0\label{v-eq} \\
\pd{t} v_{z} + S x \pd{y} v_{z} + \pd{z} p - B_{0} \pd{z} b_{z} &=& 0 \\
\pd{x} v_{x} + \pd{y} v_{y} + \pd{z} v_{z}  &=& 0\\
\pd{t} b_{x} + S x \pd{y} b_{x} - B_{0} \pd{z} v_{x} &=& 0\\
\pd{t} b_{y} + S x \pd{y} b_{y} - B_{0} \pd{z} v_{y} - S b_{x}  &=& 0\\
\pd{t} b_{z} + S x \pd{y} b_{z} - B_{0} \pd{z} v_{z} &=& 0 \label{bz-eq}
\Eeq
The divergence-free condition for the velocity implies the same for the magnetic field.
The system is 2nd order in $x$ because $\pd{x}$ only appears in two places.
All variables take the form
\Beq
p = \hat{P}(x) e^{ i ( \omega t + k_{z} z + k_{y} y ) },
\Eeq
where we note that for this section, we use the complex frequency $\omega = -i \sigma$ for the time dependence.

To make analytical progress, define the frequency ``parameters'' 
\Beq
\omega_{S}(x) \ \equiv \ \omega + S x k_{y}, \quad \omega_{A}  \ \equiv \ B_{0} k.
\Eeq
We can use equations (\ref{v-eq}--\ref{bz-eq}) to find all amplitudes in terms of $\hat{V}_{x}(x)$ and $\hat{V}_{x}'(x)$. 
\Beq
\hat{V}_{y}(x) &=& \frac{i \left(k_{y} 
   \hat{V}_{x}'(x) + k_{z}^2 \frac{ \left((f+S) \omega
   _S^2-S \omega _A^2\right)}{\omega_{S}
   \left(\omega_{S}^2-\omega _A^2\right)} \hat{V}_{x}(x)\right)}{k_{z}^2+k_{y} ^2} \\
\hat{V}_{z}(x) &=& \frac{i k_{z} \left(\hat{V}_{x}'(x)- \frac{ \left((f+S) \omega
   _S^2-S \omega _A^2\right)}{\omega_{S}
   \left(\omega_{S}^2-\omega _A^2\right)}  k_{y} 
   \hat{V}_{x}(x)\right)}{k_{z}^2+k_{y} ^2} \\
   \hat{P}(x) &=& \frac{i \left(\omega _A^2-\omega_{S}^2\right)
   \left(\hat{V}_{x}'(x) -  k_{y} \frac{ \left((f+S) \omega
   _S^2-S \omega _A^2\right)}{\omega_{S}
   \left(\omega_{S}^2-\omega _A^2\right)}
   \hat{V}_{x}(x)\right)}{\left(k_{z}^2+k_{y} ^2\right)
   \omega_{S}} \\
   \hat{B}_{x}(x) &=& \frac{\omega _A \hat{V}_{x}(x)}{\omega_{S}} \\
   \hat{B}_{y}(x) &=&\frac{\omega _A \hat{V}_{y}(x)}{\omega_{S}} -\frac{i S \omega _A \hat{V}_{x}(x)}{\omega_{S}^2} \\
   \hat{B}_{z}(x) &=& \frac{\omega _A \hat{V}_{z}(x)}{\omega_{S}}
   \Eeq
One can substitute everything into equation (\ref{u-eq}) to get a second-order equation for $\hat{V}_{x}(x)$ of the general form 
\Beq
 \hat{V}_{x}''(x) + 2 C_{1}(x) \hat{V}_{x}'(x) + C_{0}(x) \hat{V}_{x}(x) \ = \ 0,
\Eeq
where
\Beq
   C_{1}(x) \ = \  \frac{ S k_{y}  \omega _A^2}{\omega_{S} ( \omega_{S}^2-\omega _A^2 )} , \quad C_{0}(x) \ = \   \frac{\frac{f^{2} k_{z}^2 \omega_{S}^{2}}{\omega_{S}^{2}-\omega
   _A^2}+f  S  k_{z}^2 -\frac{2 S^2  k_{y}^2 \omega _A^2}{\omega_{S}^{2}}}{\omega_{S}^2-\omega _A^2}  -\left(k_{y}^{2} + k_{z}^2 \right) .
\Eeq
We can eliminate the first-order term via
\Beq
\hat{V}_{x}(x) \ = \ \chi(x)\, \psi(x) \quad \text{where} \quad \frac{\chi'(x)}{\chi(x)} \  = \ - C_{1} (x).
\Eeq
Therefore 
\Beq
-\psi''(x) + \Phi(x) \psi(x) \ = \ 0 \label{Schr-eq}, \quad \text{where}\quad
\Phi(x)  \ = \ C_{1}'(x) + C_{1} (x)^{2} - C_{0}(x).
\Eeq
Finally, by putting everything together 
\Beq
\Phi(x) \ = \ \frac{S  \left(f k_{z}^2-S k_{y}
   ^2\right) \omega _A^2 -f k_{z}^2 (f+S) \omega
   _S^2}{\left(\omega _A^2-\omega
   _S^2\right)^2} + k_{y}^{2} + k_{z}^2 .
\Eeq
The denominator of $\Phi(x)$ cannot vanish if $\omega_{S}=-i \gamma+Sxk_{y}$, with $Re(\gamma)\neq0$. 
We want to solve the Schr\"{o}dinger-type equation~(\ref{Schr-eq}) with the boundary conditions 
\Beq
\psi(x=\pm d/2) = 0.
\Eeq
The boundary conditions for $\psi(x)$ are the same as $V_{x}(x)$ because 
\Beq
\chi(x) \ \propto \ \frac{\omega_{S}(x)}{\sqrt{\omega_{S}(x)^2-\omega _A^2}} \  \ne \ 0 \quad \text{for} \quad \gamma \ \ne \ 0.
\Eeq
We solve perturbatively near the critical values for the 2D instability. We rescale each parameter in terms of a bookkeeping parameter, $\eps$.
\Beq
S  \ \to \  - \frac{\pi ^2 B_{0}^2}{d^2 f} ( 1 + \eps^{2} R), \quad \omega \ \to \ \eps^{2} \omega , \quad k_{z} \ \to \ \eps k_{z}, \quad k_{y} \ \to \ \eps^{2} k_{y},
\Eeq 
keeping in mind
\Beq
\Ro  \ \equiv  \ \frac{\pi ^2 B_{0}^2}{d^2 f^{2}}.
\Eeq
Expanding equation~(\ref{Schr-eq}) to $\mathcal{O}(\eps^{2})$, we arrive at
\Beq
\Phi(x) \ = \ -\frac{\pi^{2}}{d^{2}} - \frac{\pi^{2} \eps^{2}}{d^{2}}\left[R - \frac{d^2 }{\pi^2}k_z^2 + \frac{\Ro^2 B_0^2  k_y^2+(1+\Ro) (\omega + f \Ro\, x k_y) ^2}{ \Ro\, B_0^2
   k_z^2} \right].
\Eeq
The leading order balance is
\Beq
\psi_{0}''(x) + \frac{\pi^{2}}{d^{2}}  \psi_{0}(x) \ = \ 0  \quad \implies \quad \psi_{0}(x) \ = \ \cos(\pi x / d),
\Eeq
and next order is
\Beq
\psi_{2}''(x) + \frac{\pi^{2}}{d^{2}}  \psi_{2}(x) \ = \  \Phi_{2}(x)\psi_{0}(x) .
\Eeq
The solvability condition for $\psi_{2}(x)$ is 
\Beq
\int_{-d/2}^{d/2} \Phi_{2}(x)\psi_{0}(x)^{2} \text{d} x \ = \ 0.
\Eeq
This yields our final result
\begin{equation}\label{eq:asymp}
\sigma^{2}\ = \ - \omega^{2} \ = \  \frac{S_{c}^{2}}{1+\Ro} \frac{d^{2}}{\pi^{2}} \left[ k_{z}^{2}\left(R-\frac{d^2}{\pi^{2}}k_{z}^{2}\right)+\Upsilon\,k_{y}^{2}\right]+\ldots,
\end{equation}
where
\begin{equation}
\Upsilon \ = \  \Ro + (1+ \Ro) \frac{\pi^{2}-6}{12}  \ \approx \ 1.31\quad\text{for}\quad{\Ro} \ = \ 0.75,
\end{equation}
and we have switched back to $\sigma$ for time dependence.
The first term in equation~(\ref{eq:asymp}) results from the two-dimensional calculation.
The second term is positive definite: it \emph{always} leads to enhanced growth rates when $k_{y}\neq0$ and ultraviolet divergence in the absence of higher-order effects.

Figure~\ref{fig:growth_rate}b compares the numerical growth rates for $\SSC=1.002$ between $0\le{k_{y}}\le0.2$ to the asymptotic approximation, showing good agreement where the latter is valid.
Figure~\ref{fig:growth_rate}c shows the full spectrum for $\SSC=1.002$.
The plot shows a purely growing/decaying complex-conjugate pair (orange/blue dots on the real axis) consistent with equation~(\ref{eq:asymp}).
The other stable modes (grey dots) are rotationally modified Alfv\'{e}n waves found in left- and right-going pairs, consistent with the analytic predictions for two-dimensional stability calculations.
We reject spurious eigenvalues by using \texttt{eigentools} \footnote{\protect\url{https://github.com/dedalusproject/eigentools}} to solve equations~(\ref{eq:mhd}-\ref{eq:divu}) at two resolutions and retain only pairs equal to within one part in $10^{-6}$.

\subsection{Analysis of the transition to 2D}
\label{sec:2D_transition}

We can understand the transition from three- to two-dimensional instability semi-analytically. We do this by conducting a similar analysis as the asymptotic calculation in section~\ref{sec:asymp} near the fastest growing 2D modes, looking for values for $\SSC$ where the analogous value of $\Upsilon(\SSC)$ equals zero. 

The growth rate can be written as a function of  $k_{y},k_{z}$ and $\SSC$ near $k_{y} \approx0$.  We therefore expand the whole problem in a power-series 
\Beq
\gamma(\SSC,k_{y},k_{z}) \ = \   \sum_{n\ge 0} \gamma_{n}(\SSC,k_{z}) k_{y}^{n}, \quad \psi(x) \ = \ \sum_{n\ge 0} \psi_{n}(x) k_{y}^{n},
\Eeq
and use the same asymptotic techniques as before to obtain each progressive correction,
\Beq
\psi_{0}'' - \Phi_{0} \psi_{0} & = & 0 \label{eq:psi0} \\
\psi_{1}'' - \Phi_{0} \psi_{1} & = & \Phi_{1} \psi_{0} \label{eq:psi1} \\
\psi_{2}'' - \Phi_{0} \psi_{2} & = & \Phi_{2} \psi_{0} + \Phi_{1} \psi_{1}.
\label{eq:psi2}  
\Eeq
At leading order, we making use of the fact that all non-constant coefficients vanish for 2D modes. The leading-order eigenfunction is $\psi_{0} = \cos(\pi x / d)$, just as before. The following dispersion relations determine the maximum growth rate and the corresponding wavenumber,
\Beq
\mathfrak{D}_{0}(k_{z}) &=& \frac{f k_z^2 \left(B_0^2 S k_z^2+\gamma _0^2 (f+S)\right)}{\left(B_0^2 k_z^2+\gamma _0^2\right){}^2}+\frac{\pi ^2}{d^2}+k_z^2 \ = \ 0 \\
\frac{\mathfrak{D}_{0}'(k_{z})}{2k_{z}} &=& \frac{\gamma _0^2 f \left(B_0^2 k_{z}^2 (S-f)+\gamma _0^2 (f+S)\right)}{\left(B_0^2 k_{z}^2+\gamma _0^2\right)^3}+1 \ = \ 0.
\Eeq
We transform these relations into polynomials in terms of non-dimensional variables,
\Beq
\label{dispersion}
\y^2 \s^2 \Ro^2 +  \y \, \s  \, ( \y \, \s \, \Ro^2 + (2  - \s)  \, \Ro + \s ) \x + ( 2 \y \,\s \,  \Ro - \s  + 1 ) \x^2 +\x^3 &=& 0 \\ 
\y \, \s  \, (\y \, \s \, \Ro^2 + (2  - \s)  \, \Ro + \s    )  + 2 ( 2 \y \,\s \,  \Ro - \s  + 1 ) \x +3\x^2 &=& 0,
\label{critical-dispersion}
\Eeq
where
\Beq
\x  \ = \ \frac{d^2 k_z^2}{\pi ^2}, \quad  \quad \y \ = \ \frac{\gamma_{0}^{2}}{S^{2}}, \quad \quad \s  \ = \ \frac{S}{S_{c}}.
\Eeq
These can be computed from the parameters reported in figure~\ref{fig:growth_rate} and table~\ref{tab:runs}.
The two polynomials are straightforward to solve numerically for $\x,\y$ given $\s, \Ro$. However, we want to find critical values for $\s$, where the fastest growing modes are two dimensional. 

We find the higher-order $\gamma_{n}(\s,k_{z})$ coefficients by applying order-by-order solvability conditions. We first find that $\gamma_{1}(\s, k_{z}) = 0$ identically. This renders the first-order correction \eq{eq:psi1} solvable. The 1st-order eigenfunction takes the form 
\Beq
\psi_{1}(x) \ = \ A_{1} F(\pi x/d) \quad \text{where} \quad  F(\vartheta) = \vartheta (\vartheta \sin (\vartheta)+\cos (\vartheta))-\frac{\pi^{2}}{4} \sin (\vartheta).
\Eeq
and
\Beq
A_{1} \ = \ \frac{i \gamma _0 d^3 f S k_z^2 \left(B_0^2 (f-S) k_z^2-\gamma _0^2
   (f+S)\right)}{2 \pi ^3 \left(B_0^2 k_z^2+\gamma _0^2\right){}^{\!3}}.
\Eeq
The structure of the growth rate is therefore 
\Beq
\gamma(\s,k_{y},k_{z}) \ = \ \gamma_{0}(\s,\x) + \gamma_{2}(\s,\x) k_{y}^{2} \ + \ \ldots 
\Eeq
This is the same structure as the asymptotic calculation in section~\ref{sec:asymp}. Here, $\Upsilon \propto \gamma_{2}$ effectively. The case with dominant 2D modes corresponds to $\gamma_{2}(\s,\x)=0$.  The final result derives from the 2nd-order solvability condition 
\Beq
\int_{-d/2}^{d/2} ( \Phi_{2} \psi_{0} + \Phi_{1} \psi_{1} ) \psi_{0} \dd{x}  \ = \ 0,
\Eeq
which produces the degree-6 polynomial expression  
\Beq
& \y^{6} \s^6\Ro^6  \ + \
 \y^4 \s^5 \Ro^4   \left( (3 \xi  \s+6 \y-1)\Ro -3 \xi  \s \right) \, \x +
  \nonumber \\  
& \frac{ \y^3 \s^4 \Ro^2  }{4}  \left( (60 \y -4 \xi  (\s-8) \s -16 + 3 \s^2)\Ro^{2} + 2 \s (4 \xi  (\s+1)-3 \s) \Ro - (4 \xi -3 ) \s^2 \right) \, \x^{2} + \nonumber \\ 
& \frac{\y^2  \s^3 \Ro }{2}  \left( (3 \s^2-4 \xi  (\s-3) \s+40 \y-12) \Ro^{2} + 24 \xi \s \Ro + (4 \xi -3) s^2 \right) \, \x^{3} + 
\nonumber \\ 
&  \frac{\y \,\s^2}{4} \left( (60 \y -(4 \xi -3) \s^2 - 16)\Ro^{2} + 2\s(3 \s-4 \xi  (\s-3)) \Ro - (4 \xi -3) \s^2 \right)\, \x^{4} +\nonumber \\ 
 &  \s\, (( 6 \y -\xi  \s-1) \Ro- \xi  \s) \, \x^{5}  \ + \   
   \x^{6} \ = \ 0.
   \label{2D-condition}
\Eeq
The new coefficient is
\Beq
\xi \ = \ \frac{\pi^{2}-6}{12} \ = \ \frac{\int_{-d/2}^{d/2} x^2 \cos ^2\left(\frac{\pi  x}{d}\right) \dd{x}}{\frac{d^{2}}{\pi^{2}}\int_{-d/2}^{d/2}  \cos ^2\left(\frac{\pi  x}{d}\right) \dd{x}} \ \approx \ 0.322467.
\Eeq
This results from projecting the squared shear onto the leading-order eigenmodes, just as in the previous section. 

We solve \eq{dispersion}, \eq{critical-dispersion}, and  \eq{2D-condition} for $\x,\y,\s$ using Newton's method for a variety of $\Ro$, yielding

\begin{center}
\begin{tabular}{cccccccccccccc}
\textbf{$\Ro$} & \textbf{$k_{z} d/\pi$}  & \textbf{$\gamma_{\max}/|S_{2D}|$}  &  \textbf{$S_{2D}/S_{c}$} \\
                   1.00  & 0.689610 & 0.208727 &   2.09694 \\ 
                   0.75 & 0.682819 & 0.214895 & 2.05136 \\
                   0.50 & 0.676062 & 0.221309 & 2.00061 \\
                   0.25 & 0.671528 & 0.228551 & 1.94963 \\
                   0.10 & 0.673164 & 0.234376 & 1.92692 \\
                   0.00 & 0.462324 & 0.240212 & 1.92465
\end{tabular}
\end{center}
The value of $\SSC = 2.05136$ at the 2D to 3D transition when $\Ro = 0.75$ matches the numerical results shown in Figure~\ref{fig:growth_rate} two all reported significant figures in the numerical calculation. 

We conclude this section with some speculations concerning the higher order corrections. Because of symmetry, $\gamma_{3}(\s,\x) = 0$ identically. This means that to leading order 
\Beq
\gamma(\s,k_{y},k_{z}) \ = \ \gamma_{0}(\s,\x) + \gamma_{2}(\s,\x) k_{y}^{2}  + \gamma_{4}(\s,\x) k_{y}^{4} \ + \ \ldots
\Eeq 
For $\SSC > S_{2D}/S_{c}$, $\gamma_{2} > 0$. This is sufficient to determine the most unstable 3D wavenumber near the transition to dominant 2D modes, 
\Beq
k_{y}^{2} \ \approx \ -\frac{\gamma _2}{2 \gamma_4} \  \propto \   S_{2D}/S_{c} - \SSC   . 
\Eeq 
In general, $\gamma_{4} \ne 0$. We speculate that $\gamma_{4} < 0$, though the calculation is outside the scope of this work. If there is a negative 4th-order feedback, then 
\Beq
k_{y} \ \sim \ \sqrt{S_{2D}/S_{c} - \SSC}, 
\Eeq
which is consistent with Figure~\ref{fig:growth_rate}d near the critical point. However, a scenario where $\gamma_{4} > 0$ would require behaviour that is inconsistent with our numerical calculations.

\subsection{Consistency with previous simulations}
\label{sec:prior_work}

Numerical simulations of the MRI in accretion disks consistently show axisymmetric modes dominating the early evolution of the MRI before breaking down into 3D MHD turbulence \cite{1995ApJ...440..742H,2018ApJ...853..174H,2019ApJS..241...26D}. 
These ``channel modes'' are exact non-linear solutions for $x$-shearing-periodic domains, and can only saturate via parasitic shear instabilities \cite{1994ApJ...432..213G}.
Impenetrable and stress-free boundary conditions are more applicable to stellar interiors. 
But even in disks, any kind of finite radial extent will cutoff the unbounded growth of channel modes. 
We reconcile our results with earlier simulations by showing that the MRI indeed prefers two dimensions for the larger values of criticality found in disk simulations. 
Figure~\ref{fig:growth_rate}d shows the phase angle $\phi=\arctan(k_{y}/k_{z})$ of the fastest growing mode as function of $\SSC$.
The overall critical shear for three-dimensional modes is $S_{c,\text{3D}}\simeq0.102S_c$.
Above $\SSC\gtrsim2.05$, $\phi$ becomes zero, indicating that axisymmetric modes have the fastest growth rates.
For the fiducial run in \cite{1996ApJ...464..690H}, $\SSC\simeq4.84$ (most works use similar values).
Thus, our theory predicts axisymmetric modes should dominate the linear dynamics for parameters studied in prior numerical simulations.
However, it may be that those simulations saw a predominance of axisymmetric modes because of the radial boundary conditions and not their large shear rates.

Even for $\SSC> 2.05$ there are significant swaths of unstable three-dimensional modes with growth rates comparable to the maximum (figure~\ref{fig:growth_rate}a).
Also, non-normality generically accompanies non-axisymmetry in shearing systems \cite{1992MNRAS.255P..25K}.
In non-normal systems, transient amplification can occur for stable modes, and can even cause turbulence.
Therefore significant three dimensionality likely implies important non-normal behaviour near onset.

\subsection{Eigenvectors}
\label{sec:eigenvectors}

\begin{figure}[h!]
  \centering
  \includegraphics[width=\columnwidth]{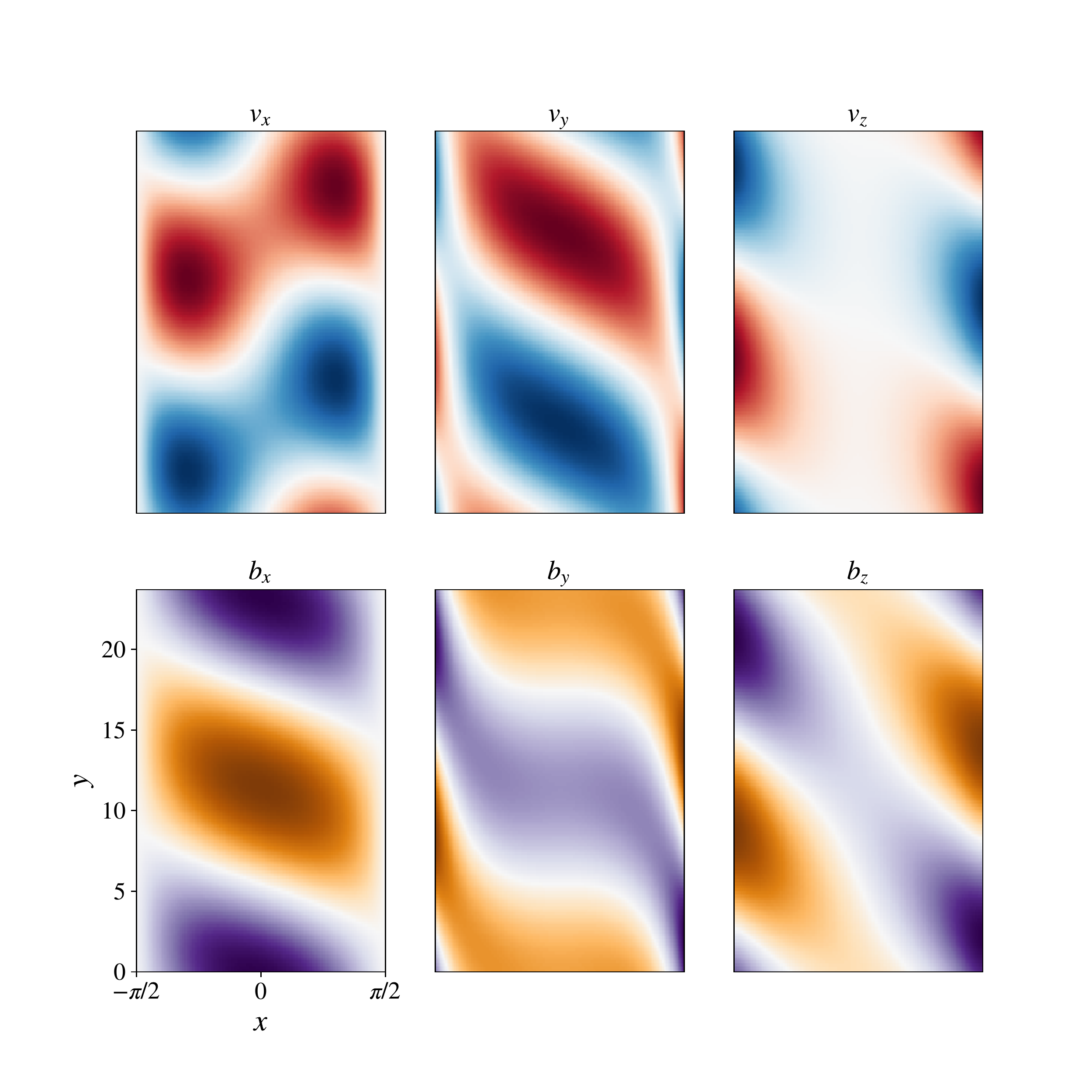}
  \caption{\textit{Ideal} eigenvectors of the velocity and magnetic field perturbations for the most unstable mode when $\SSC=1.02$, and $\eta=\nu=0$; calculations at $\eta=\nu=10^{-5}$ are indistinguishable.
  The top row shows $v_{x}$, $v_{y}$, $v_{z}$ (red is negative, blue positive); the bottom row shows $b_{x}$, $b_{y}$, $b_{z}$ (purple is negative, orange positive). 
The amplitudes are arbitrary because of linearity.}
  \label{fig:eigvec}
\end{figure}

Figure~\ref{fig:eigvec} shows the eigenvectors for the most unstable mode at $\SSC=1.02$ using ideal MHD. 
No critical layers can form for $\gamma\ne0$.
We therefore solve equations~(\ref{eq:mhd}--\ref{eq:divu}) without dissipation as a 2nd-order system in $\partial_{x}$ and impose $v_{x}=0$ at $x=\pm{d/2}$.
The eigenfunctions are indistinguishable from those at finite dissipation, lending additional confidence to our other results.
The tilted structures of $v_{x}$ and $v_{y}$ as well as $b_{x}$ and $b_{y}$, imply non-trivial Reynolds and Maxwell stresses.

\subsection{Reynolds Number and Departure from Ideal MHD}
\label{sec:reyn}

In order to ensure that our calculations are probing the ideal MHD regime we are interested in, we have performed a series of runs (24-27) with $\eta = \nu$ ranging from $10^{-6}$ to $10^{-2}$, equivalent to varying the Reynolds and Magnetic Reynolds numbers $\Reyn = \Reym = 10^2$ -- $10^6$.
\begin{figure}[h!]
  \centering
  \includegraphics[width=\textwidth]{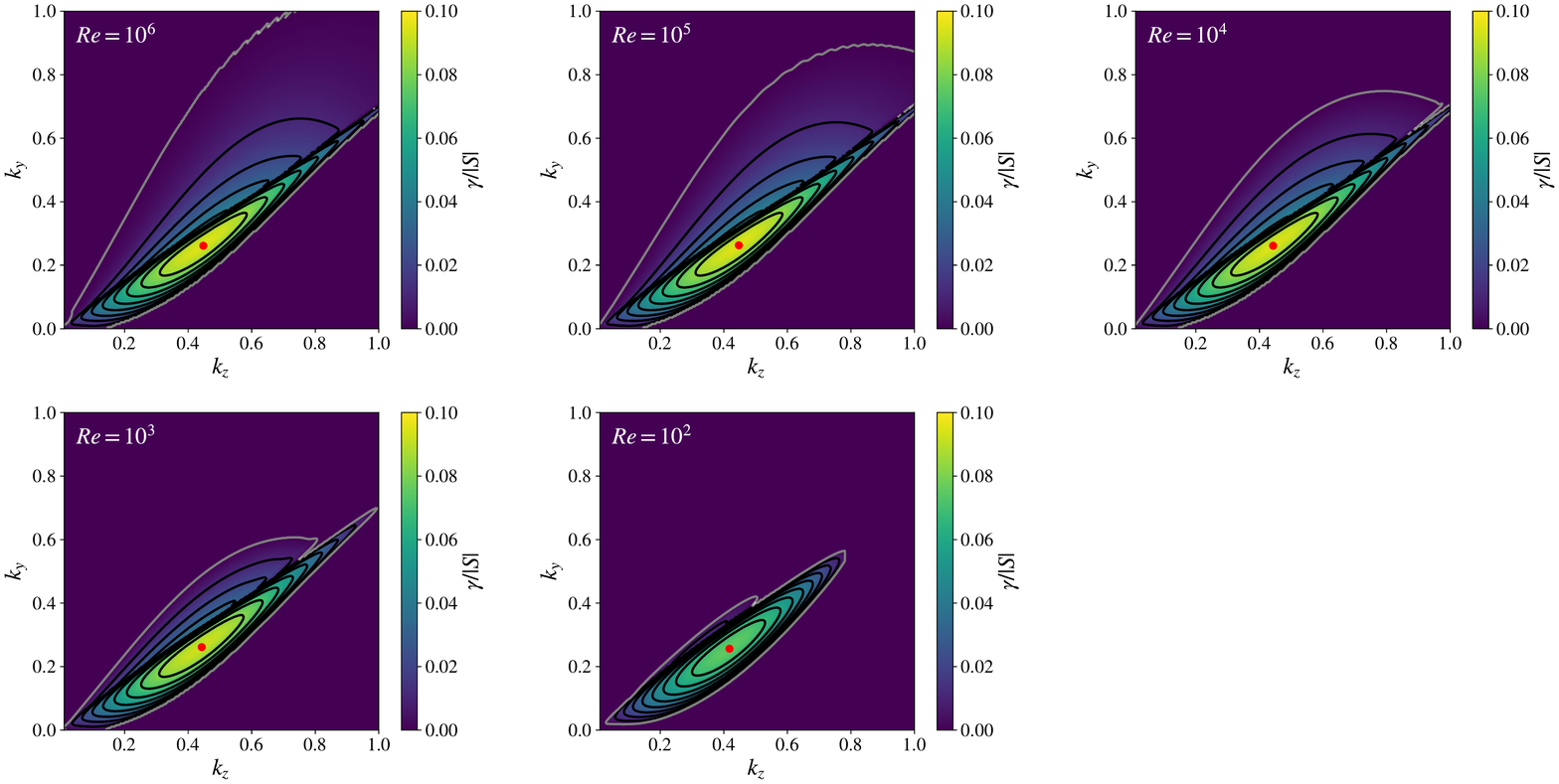}
  \caption{Growth rates at $\SSC = 1.02$ for varying diffusivities. Note that the top three panels are essentially identical, suggesting that above $\Reyn \gtrsim 10^4$ we are reasonably close to ideal conditions.}
  \label{fig:reynolds}
\end{figure}
Figure~\pageref{fig:reynolds} shows the growth rates at $\SSC=1.05$ for five values of $\Reyn = \Reym$.
For $\Reyn \gtrsim 10^4$, the maximum growth rate is located at the same position and the same growth rate; the unstable region expands, as expected.
This suggests that our results at $\Reyn = 10^5$ are reasonably close to ideal MHD for the purposes of understanding the MRI's preference for 3D modes near shear onset.

\subsection{Mean electromotive forces}
\label{sec:emf}
We should expect the generation of a mean electromotive force (EMF) in the linear regime, \textit{i.e}
$\left<\boldsymbol{v}\times\boldsymbol{b}\right>\sim\,\alpha\!\left<\boldsymbol{B}_{0}\right>$.
This is significant because of the possibility of direct laminar dynamo action in regions of weak shear at large Reynolds number.
\begin{figure}[h!]
  \includegraphics[width=\columnwidth]{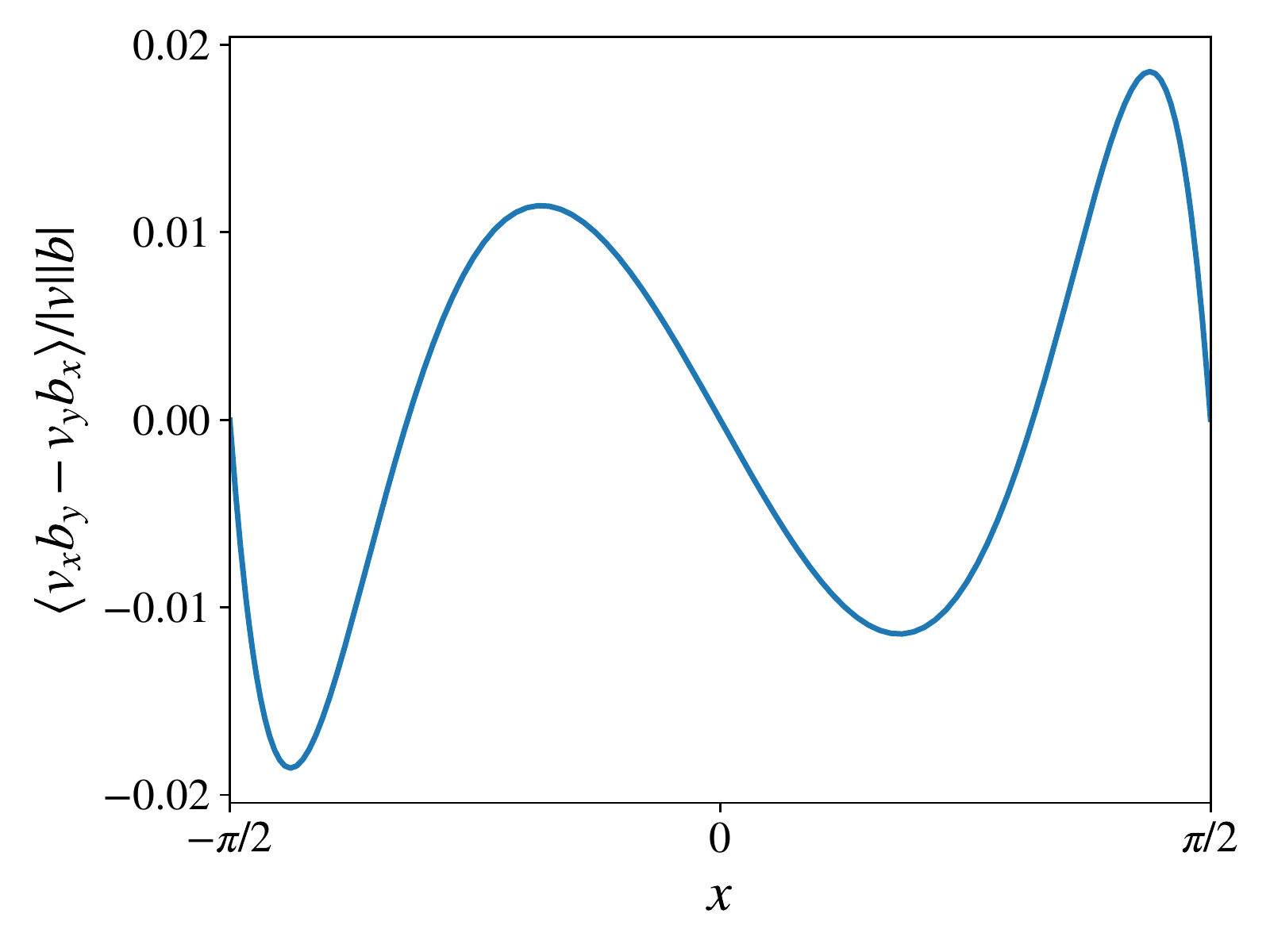}
  \caption{Velocity-magnetic field correlations relevant to the mean electromotive force for the eigenvectors of the fastest growing mode at $\SSC = 1.02$.}
  \label{fig:correlation}
\end{figure}
Figure~\ref{fig:correlation} shows the correlation (averaged over $y,\,z$) as a function of $x$ over the domain for $\SSC = 1.02$,
\begin{equation}
\text{EMF}\propto\frac{\left<\boldsymbol{\hat{z}}\cdot(\boldsymbol{v}\times\boldsymbol{b})\right>_{\!y,z}}{\boldsymbol{v}_{\text{rms}}\boldsymbol{b}_{\text{rms}}},
\end{equation}
where ``rms'' is denotes the average over the whole domain. 
The correlation is independent of the arbitrary normalisation of the linear eigenvectors.
The non-zero correlation shows the tendency for back reaction on the mean magnetic field.

MRI dynamos have been studied in a number of different contexts \cite{2007PhRvL..98y4502R,2011ApJ...740...18O,2015PhRvL.114h5002S}, but all except one focused on non-linear, usually turbulent, dynamos. 
The only exception we are aware of, \cite{2016MNRAS.462..818B}, found exponential growth of mean magnetic fields during the linear growth phase of the MRI far from stability in numerical simulations.
Our work explains this result in terms of purely linear dynamics:
non-axisymmetric MRI unstable modes drive the dynamo growth of magnetic fields.

\subsection{Extensions to stellar and experimental conditions}
\label{sec:extensions}

Finally, we vary two important parameters:
$\Ro$ and the magnetic Prandtl number $\Prm=\nu/\eta$.
\begin{figure}[h!]
  \centering
  \includegraphics[width=\columnwidth]{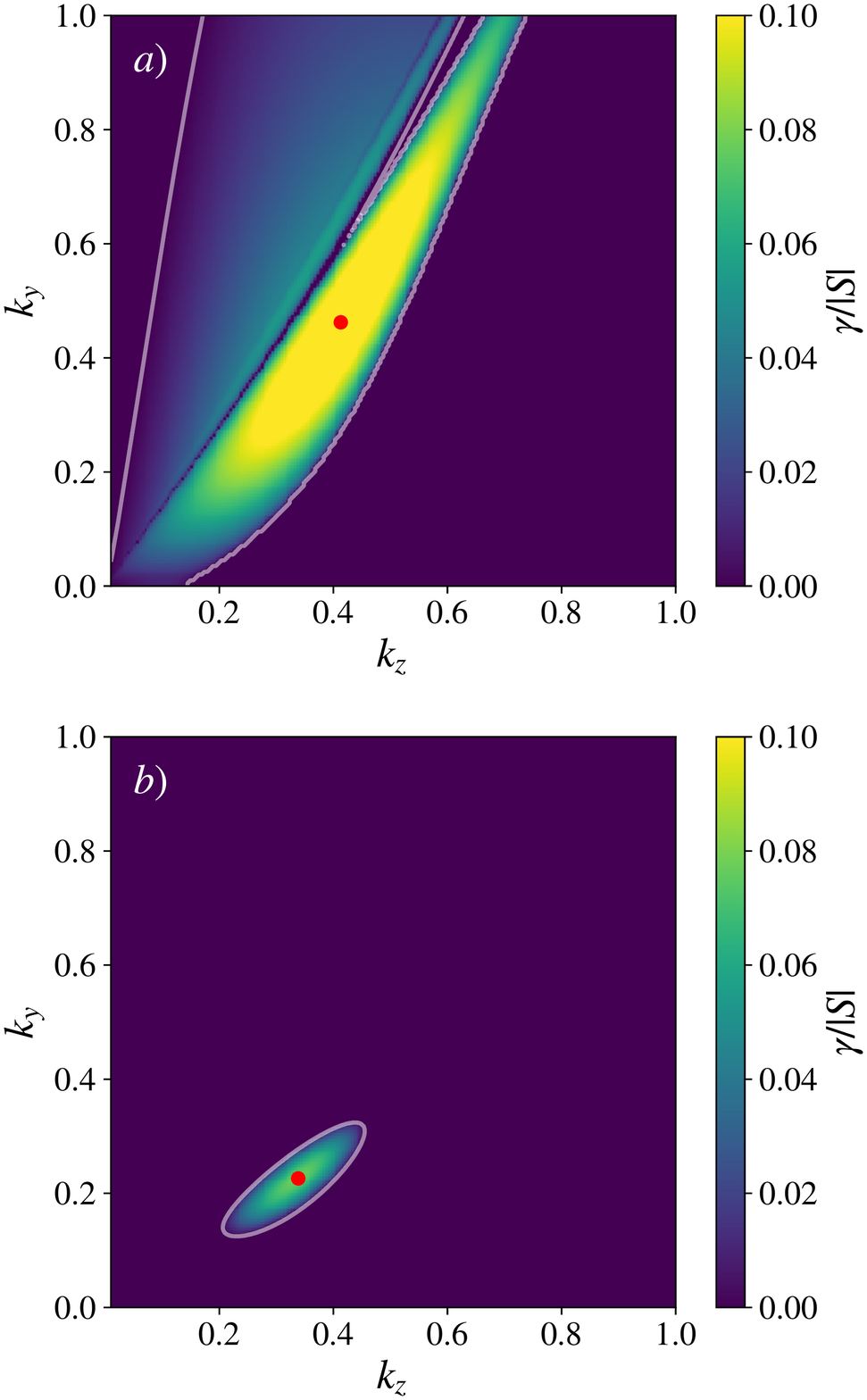}
  \caption{(a) Growth rates for $\Ro=1/10$ relevant to the stellar interiors. 
  (b) Growth rates for MRI near onset at liquid metal-like diffusivities, $\nu=10^{-6}$, $\eta=0.1$.
The lack of two dimensional instability is evident by the fact that the zero contour does not intercept the x-axis.}
  \label{fig:other_params}
\end{figure}
Figure~\ref{fig:other_params} shows the growth rates for $\SSC=1.02$.
The upper panel shows $\Ro=1/10$ (addressing a possible range within stellar interiors) holding all other parameters equal to their fiducial values.
The growth rates show similar behaviour to the $\Ro=3/4$ cases, \textit{i.e.}, dominated by three-dimensional modes with $\gamma_{\max}/|S|\simeq0.1$.
The lower panel shows $\nu=10^{-6}$ and $\eta=0.1$ with all other parameters equal to their fiducial values. 
This case is relevant to liquid metals e.g., the Princeton MRI experiment \cite{2002JFM...462..365G}
at $\Prm=10^{-5}$. 
Our results are for the rotating plane Couette geometry (Taylor-Couette in the small-gap limit $R_{2}\approx{R}_{1}$) with highly idealised boundary conditions.
It is nevertheless quite interesting that, near onset, the low-$\Prm$ MRI is \emph{only} unstable to three-dimensional modes.

\section{Conclusion}
\label{sec:conclusion}

Our results show that three-dimensional modes grow faster than two-dimensional modes whenever the MRI is near its critical shear values.
While we mainly focus on the ideal Keplerian case ($\Ro=3/4$), we also demonstrate robustness for low Rossby ($\Ro=1/10$) and magnetic Prandtl numbers ($\Prm=10^{-5}$).
There are several important future directions this work suggests.
First, the Sun possesses two internal shear layers with inwardly increasing shear, the high-latitude tachocline, and the near-surface shear layer (NSSL). 
Past work has already pointed out the possibility of the small-scale MRI in the Sun using local analysis \cite{2007ApJ...667L.207P,2011MNRAS.411L..26M,2014ApJ...787...21K}.
The NSSL, in particular, may also host slower MRI-driven dynamics, despite containing small-scale convection.
The NSSL contains the strongest shear anywhere within the solar interior;  $\Omega(r) \ \propto \ 1/r$, implying $\Ro = 1/2$.
It is therefore crucial to further elucidate the nonlinear saturation of the 3D MRI, along with its robustness to convection.
An important uncertainty with regard to stellar applications is the fact that $\SSC$ depends on the square of the magnetic field strength, a quantity that is rather uncertain.
Second, our low-$\Prm$ results suggest that three-dimensional MRI modes may be the easiest to excite in liquid metal experiments.
Determining possible non-axisymmetric signatures in Taylor-Couette experiments requires followup work using more realistic boundary conditions and geometry. 
Our prior work on the axisymmetric MRI \cite{2017ApJ...841....1C,2017ApJ...841....2C} shows only small differences between rotating plane and cylindrical Taylor-Couette geometries. 
We fully expect the general three-dimensional features of the MRI to persist in more complex applications.


\dataccess{All code used for this project is available at \url{https://github.com/jsoishi/mri\_prefers\_3D}}

\aucontribute{J\@.S\@.O and G\@.M.\@V\@. led the project.
J\@.S\@.O prepared all figures.
G\@.M\@.V performed the asymptotic calculation.
M.B.\ ran the bulk of the eigenvalue calculations and prepared several preliminary figures.
A\@.S\@. performed preliminary eigenvalue calculations which led to the fiducial parameter choices used here.
J\@.S\@.O, G\@.M.V\@., K\@.J\@.B., D\@.L\@., and B.P\@.B developed \emph{Dedalus}, which is used for all calculations in this paper.
All authors contributed to the writing of the manuscript and physical interpretation of the results.
}

\competing{We declare we have no competing interests.}

\funding{Oishi, Baxter, and Brown acknowledge support from NASA LWS grant No. NNX16AC92G.
Oishi also acknowledges support from Research Corporation Scialog Collaborative Award (TDA) ID \#24231.}

\ack{Computations were performed on the \emph{Leavitt} cluster at the Bates College High Performance Computing Centre.}


\bibliographystyle{RS}
\bibliography{mri}

\end{document}